\documentclass[letterpaper,english,aps,prl,twocolumn,floatfix,showpacs,amsfonts,amssymb,superscriptaddress]{revtex4}

\usepackage[T1]{fontenc}
\usepackage[latin1]{inputenc}
\usepackage{graphics}
\usepackage{amssymb}
\usepackage{overpic}

\newcommand{\beq}{\begin{equation}}
\newcommand{\eeq}{\end{equation}}
\newcommand{\bea}{\begin{eqnarray}}
\newcommand{\eea}{\end{eqnarray}}

\newcommand{\urhcoge}{URh$_{1-x}$Co$_x$Ge}
\newcommand{\urhgesi}{URhGe$_{1-x}$Si$_x$}
\newcommand{\urhge}{URhGe}
\newcommand{\ucoge}{UCoGe}

\begin{document}

\title{Hybridization driven quantum critical behavior in weakly-itinerant ferromagnets}
%\title{Ferromagnetism in nearly critical, strongly hybridized ternary uranium compounds}

\author{M.~B.~Silva~Neto}

%\email{mbsn@if.ufrj.br}

\affiliation{Instituto de F\' isica, Universidade Federal do
Rio de Janeiro, Caixa Postal 68528, Rio de Janeiro, Brasil}

\author{A.~H.~Castro~Neto}

%\email{neto@bu.edu}

\affiliation
{Department of Physics, Boston University, Boston, MA 02215, USA}

\affiliation
{Graphene Research Centre and Department of Physics, 2 Science Dr. 3, National
University of Singapore, 117542.}

\author{J.~S.~Kim}

%\email{kim@phys.ufl.edu}

\affiliation
{Department of Physics, University of Florida, Gainesville, FL 32611-8440, USA}

\author{G.~R.~Stewart}

%\email{stewart@phys.ufl.edu}

\affiliation
{Department of Physics, University of Florida, Gainesville, FL 32611-8440, USA}

\date{\today}

\received{in ...}

\begin{abstract}

We investigate the unusual magnetic properties of nearly-critical, weakly-itinerant 
ferromagnets with general formula UTX, where T=Rh,Co and X=Ge,Si. 
As a unique feature about these systems, we show that changes in the 
$V_{df}$ hybridization control their proximity to a ferromagnetic instability, 
and determine the evolution of: the ground state magnetization, $M_0$, the 
Curie Temperature, $T_C$, the density of states at the Fermi level, 
$N(E_F)$, the $T^2$ resistivity coefficient, $A$, and the specific heat coefficient, 
$\gamma$. The universal aspect of our findings comes from the dependence
on only two parameters: the T$_d$ bandwidth, $W_d$, and the distance 
between T$_d$ and U$_f$ band centers, $C_{T_d}-C_{U_f}$.

%We describe the mechanism behind the unusual evolution of the ground state 
%magnetization, $M_0(x)$, and Curie temperature, $T_C(x)$, as a function of 
%isovalent-disorder, $x$, in strongly-hybridized, nearly critical, weakly-itinerant 
%ferromagnets with formula UTX, where T=Rh,Co and X=Ge,Si. 
%We show that changes in the $V_{df}$ hybridization control the proximity of the 
%system to a ferromagnetic instability and determine the evolution of not only
%$M_0(x)$ and $T_C(x)$, but of other quantities such as the density of states at 
%the Fermi level, $N(E_F)$, the $T^2$ resistivity coefficient, $A$, and the 
%specific heat coefficient, $\gamma$. Using only two parameters: the 
%T$_d$ bandwidth, $W_d$, and the distance between T$_d$ and U$_f$ band 
%centers, $C_{T_d}-C_{U_f}$, we obtain good agreement with experiments in {\urhcoge}.

\end{abstract}

\maketitle

In the past few years, a new class of materials has attracted significant 
attention due to their ability to sustain coexisting ferromagnetism and 
superconductivity at ambient pressure. These are ternary U compounds of 
general formula UTX, where T$=$ Rh, Co, and X=Ge  \cite{Aoki,Huy}. 
Together with UGe$_2$, where superconductivity with $T_c=0.8$ K is 
obtained under applied pressure in a regime where ferromagnetism is
still robust, $T_C=35$ K \cite{Saxena}, and possibly UIr \cite{UIr}, 
these materials have joined the special family of spin-triplet, 
weakly-itinerant ferromagnetic superconductors. Furthermore, the 
observation that both superconductivity and ferromagnetism are very 
susceptible to applied pressure, indicates that the Cooper pairing is 
being induced by the proximity to a ferromagnetic quantum critical 
point (QCP) \cite{Pfleiderer}.

The search for a ferromagnetic QCP has thus encountered a natural 
candidate, {\urhge}.
%, because it is already relatively close 
%to a ferromagnetic  instability, with $T_C=9.5$ K, at ambient pressure 
%\cite{Aoki}. 
Here the QCP could be reached through the chemical substitution of 
the transition metal element, Rh, since other isostructural compounds 
are either paramagnets, as URuGe, or even closer to the ferromagnetic 
instability, as {\ucoge}. Thus it seemed natural to investigate the evolution 
of the ferromagnetism and/or superconductivity in URh$_{1-x}$T$_{x}$Ge, 
for T$=$Ru or Co \cite{Sakarya}. This was considered by Sakarya 
{\it et al.} in \cite{Sakarya}, but, although the suppression of the ferromagnetism 
went according to expectations in the case of Ru, at $x_c=0.38$, for 
the case of Co, instead, a huge and unexpected enhancement of 
$T_C$ (larger than $100\%$) was observed in $T_C$, before it dropped 
down to less than $3$ K, as $x\rightarrow 1$. The unexpected aspect of 
this finding comes from the fact that the isovalent doping in {\urhcoge} 
does not introduce or remove carriers and the enhancement of 
$T_C$ is much larger than what would be expected from volume 
compressibility effects (less than $30\%$) \cite{Sakarya}. 

In this letter we describe the mechanism behind such unusual behavior 
of $T_C$ as a function of isovalent disorder, $x$, in {\urhcoge}, and 
we explain why a similar behavior is not observed in  {\urhgesi}. We show 
that the evolution of $T_C$(x) is governed by changes in the hybridization,
$V_{df}$,  between the transition metal T$=$Rh,Co $d$-band and the 
actinide U $f$-level, which influences also other quantities such as density 
of states at the Fermi level, $N(E_F)$, the $T^2$ resistivity coefficient, 
$A$, and the specific heat coefficient, $\gamma$. As we shall demonstrate,
hybridization ultimately controls the proximity to the ferromagnetic instability, 
or QCP, where marked changes of behavior are observed for different ground 
state properties in such weakly-itinerant ferromagnetic systems.

The ternary {\urhge} and {\ucoge} compounds crystalize in the TiNiSi 
orthorhombic crystal structure with $P(nma)$ space group. There are 
$4$ chemical formula units per unit cell and each of the $4$ U, Rh/Co, 
or Ge atoms occupy one of the $4(c)$ crystallographic positions $(x,1/4,z)$,
$(1/2-x,3/4,1/2+z)$, $(1/2+x,1/4,1/2-z)$, and $( -x,3/4,-z)$,
resulting in a primitive lattice. The relative positions of the U,
Rh/Co and Ge atoms in the unit cell were determined by X-ray
($300$ K) and neutrons ($20$ K) \cite{Prokes}:
$x_U=0.9959(9)$, $z_U=0.2038(05)$, $x_{Rh/Co}=0.5716(7)$,
$z_{Rh/Co}=0.5730(17)$, $x_{Ge}=0.6103(9)$, $z_{Ge}=0.5898(10)$,
each of these quantities being measured in units of the lattice
parameters $a=6.8552$ {\AA}, $b=4.3274$ {\AA}, and $c=7.5010$
{\AA} for {\urhge}, and $a=6.8450$ {\AA}, $b=4.2060$ {\AA}, and
$c=7.2220$ {\AA} for {\ucoge}.
Samples of {\urhcoge} with $x=0.0$, $0.2$, $0.3$, $0.4$, $0.5$,
$0.6$, $0.7$, $0.8$, and $0.9$ were prepared by arc-melting the
constituent elements (depleted U from Los Alamos National
Laboratory, $99.9+\%$ purity, Rh/Co/Ge from Johnson Matthey Aesar,
$99.95\%/99.995\%/99.9999\%$ purity) under a purified Ar
atmosphere. Samples were x-rayed using powder diffractometry in
the University of Florida Major Analysis Instrumentation Center.

%
%%%%%%%%%%%%%%%%%%%%%%%%%%%%%%%%%%%%%%%%%%%%%%%%%%%%%%%%%%%
\begin{figure}[h]
\includegraphics[scale=0.3]{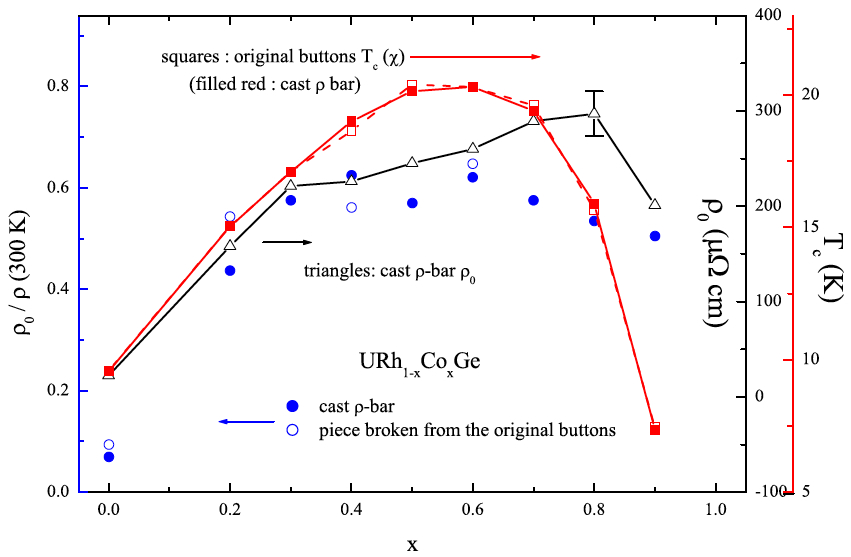}
\caption{(Color online) Experimental data for $T_C$ as a function
of doping in {\urhcoge} (right vertical axis), as well as resistivity 
data (left vertical axis).}
\label{Exp-Phase-diag}
\end{figure}
%%%%%%%%%%%%%%%%%%%%%%%%%%%%%%%%%%%%%%%%%%%%%%%%%%%%%%%%%%
%

In a previous work on the effect of disorder on $T_C$ 
\cite{ucusige-prl}, we found that sizable variations 
in the electronic mean free path, $\ell$, could drive 
the system from a clean towards a diffusive regime, 
where $T_C$ was found to be enhanced. We thus wanted 
to compare the evolution of $T_C(x)$ with the variation, 
with $x$, of disorder in {\urhcoge} and {\urhgesi}, 
using, as straightforward measures of the disorder, 
the residual resistivity $\rho_0$ or the inverse of 
the residual resistivity ratio, $\rho_0/\rho(300 \mbox{K})$. 
%We immediately discovered 
%that arc-melted buttons of {\urhcoge}
%contained voids and cracks, making it impossible to accurately
%measure the residual resistivity. Although one usually in such a
%situation then simply measures the ratio (as was done in our
%previous work \cite{ucusige-prl}) of the resistances to obtain an
%indicator of disorder, there is another approach which we
%have used (Ref. \cite{ucusige-prl}) successfully to obtain $\rho(0)$.  
%Resistivity bars of known geometrical cross section can be made by 
%sucking a molten bead of each composition $x$ of {\urhcoge} down 
%into a water-cooled rectangular hole of known cross section. These 
%{\it sucker bars}, as has been found in the past \cite{ucusige-prl}, 
%avoid the problems of cracks and voids when measuring the resistivity.  
%Although the agreement between the values for $\rho(0)/\rho(300 K)$ 
%for the sucker bars and the as-prepared material is not as perfect 
%(see Fig.\ \ref{Exp-Phase-diag}) as for the $T_C$ values between 
%the two kinds of material, the trend is the same in both the as-melted 
%and sucker bar samples. 
As shown in Fig.\ \ref{Exp-Phase-diag}, the Curie temperatures 
measured via magnetic susceptibility for the as-melted chunks 
and the resistivity bars (formed by sucking the molten material 
into a water-cooled mold, see \cite{ucusige-prl}) agree well. 
However, as Fig.\ \ref{Exp-Phase-diag} also shows, the measures 
of disorder (either $\rho(0)$ or $\rho(0)/\rho(300 K))$ did not 
appear to vary as $T_C(x)$, where the peak at $x=0.6$ is quite 
marked. The conclusion is that, for {\urhcoge}, the electronic 
mean free path, $\ell$, is always large enough, $k_F\ell\gg 1$, 
such that the system is always in the clean limit, and thus the 
variation in $T_C$ must be controlled by a different mechanism.

The magnetic properties of ternary actinide compounds with general 
formula UT(Si,Ge), are determined by the $5f$-$nd$-hybridization, 
$V_{df}$, with $n=3,4,5$ \cite{Gasche}. Quite generally, as one goes down 
a fixed column in the periodic table (isovalent elements), the $3d,4d$ 
and $5d$ bands of the transition metals become broader, as well as 
lower in energy with respect to the position of the $f$-level of the U 
ion \cite{Gasche}. This is in fact observed for the UCoAl, URhAl, UIrAl
series \cite{Gasche}, as well as for the UCoGe and URhGe compounds
here under study \cite{Divis-Rh,Divis-Co}. As a consequence, the 
$V_{df}$ hybridization can be written as \cite{Brooks-Gloetzel}
\beq
V_{df}=\frac{W_d W_f}{C_{T_d}-C_{U_f}},
\eeq
where $W_d, W_f$ are the transition metal and $f$-level bandwidths, 
and $C_{T_d}$ and $C_{U_f}$ are the centers of the T$_d$ and U$_f$ 
bands respectively. In terms of the muffin-tin orbital (MTO) theory 
by Andersen \cite{Andersen}, the above formula becomes
$V_{df}=(\eta_{df} \hbar^2/m^*)\sqrt{r_d^{5} r_{f}^{7}}/d^{6}$, where
$m^*$ is the conduction band effective mass, $\eta_{df}$ are 
dimensionless angular momentum dependent coefficients, $r_d$ and $
r_f$ are the atomic radii of the $nd$ and $5f$ elements, and 
$d$ is the interatomic distance \cite{Harrison}.

\begin{figure}[h]
\includegraphics[scale=0.244]{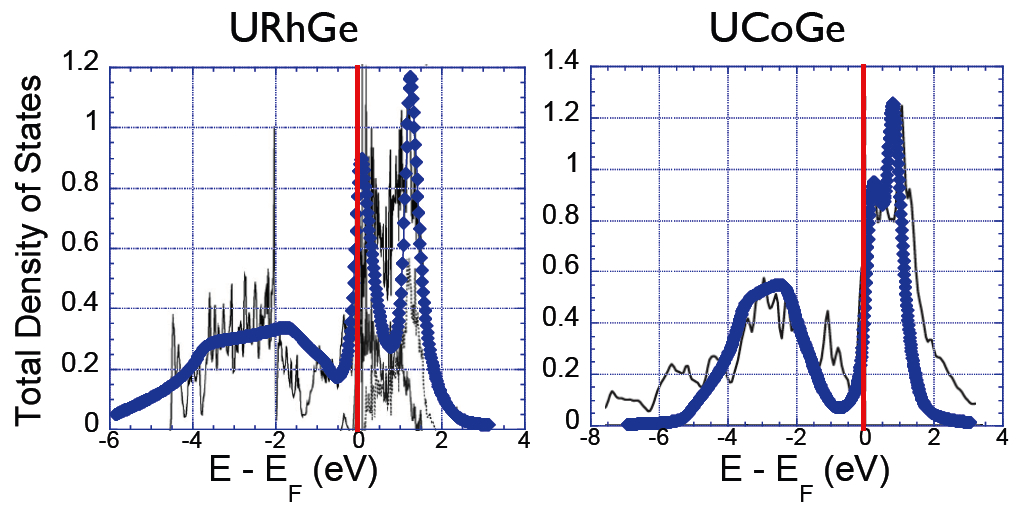}
\caption{(Color online) Total DOS for {\urhge} and 
{\ucoge}. The blue (diamond) points are numerical
data from Eqs.\ \ref{Self-Consistent-Eq} and the
solid black lines are numerical data from LSDA+U,
for URhGe \cite{Shick}, and from DFT, for UCoGe \cite{Divis-Co}.}
\label{Fig-DOS-PAM}
\end{figure}

According to density functional theory (DFT) calculations, for
{\urhge} the $4d$-band of Rh starts at about $-6$ eV, relative to the
Fermi level, and extends beyond $2$ eV above the Fermi level
\cite{Divis-Rh}, thus being very broad. Furthermore, the Fermi 
level is located in a region of high DOS with $5f$-character, 
see Fig.\ \ref{Fig-DOS-PAM}. The Stoner criterium is satisfied 
and {\urhge} exhibits ferromagnetic order with $T_C\approx 9.5$ K. 
This is also consistent with the high electronic contribution to 
the specific heat for {\urhge}, which has been estimated as 
$\gamma\approx 115$ mJ/mol K$^2$ \cite{Prokes} (after 
subtraction of the magnetic contribution). The results from DFT 
for URhGe also agree quite well with LSDA+U calculations 
\cite{Shick}. For {\ucoge}, DFT shows that the $3d$-band of Co 
starts at about $-4.5$ eV and extends up to approximately $1.6$ 
eV, being, as such, much sharper \cite{Divis-Co}. Here, instead, 
the Fermi level is located in a region of much smaller $5f$ DOS, 
see Fig.\ \ref{Fig-DOS-PAM}, and the Stoner criterium is barely 
satisfied. In this case, {\ucoge} is very close to the Stoner instability 
and is thus a nearly critical ferromagnet with $T_C\approx 3$ K. 
Again this is consistent with the smaller electronic contribution 
to the specific heat for {\ucoge}, given by $\gamma\approx 65$ 
mJ/mol K$^2$ \cite{Sakarya}. 

%
%%%%%%%%%%%%%%%%%%%%%%%%%%%%%%%%%%%%%%%%%%%%%%%%%%%%%%%%%%%
%%%%%%%%%%%%%%%%%%%%%%%%%%%%%%%%%%%%%%%%%%%%%%%%%%%%%%%%%%
%

The microscopic mechanism for the unusual behavior for the Curie 
temperature in the {\urhcoge} series can now be described. As Rh 
$\rightarrow$ Co, the $d$-bandwidth $W_d$ becomes smaller as 
it changes its character from $4d\rightarrow 3d$ (we consider $W_f$
fixed, for simplicity). The smaller $W_d$ 
reduces the hybridizaiton, $V_{df}$, which causes a large increase 
of the $f$-DOS at the Fermi level. Now, within the usual approach to 
itinerant ferromagnetism, one writes $T_C\propto(I N(E_F)-1)^{(z/z+1)}$, 
where $I$ is the so called Stoner integral and $z=3$ for a clean 
itinerant ferromagnet \cite{Moriya}. Thus, we see that an increase of $N(E_F)$ 
leads to an {\it enhancement} of $T_C$.  On the other 
hand, as doping further increases, the bottom of the $d$-band is 
shifted to higher energies, $C_{T_d}$ moves towards $C_{U_f}$, 
since the $3d$-band is closer to the $f$-level than $4d$-band. 
The closeness between the $d$ and $f$ bands increases the 
hybridization, $V_{df}$, which reduces the $5f$-DOS at the Fermi 
level, reducing $T_C$ considerably. Thus, the non-monotonic behavior 
observed experimentally in the $T_C(x)$ phase diagram of {\urhcoge} 
results from the combination of the two effects on $V_{df}$ (or $N(E_F)$) 
when taken simultaneously.

%
%%%%%%%%%%%%%%%%%%%%%%%%%%%%%%%%%%%%%%%%%%%%%%%%%%%%%%%%%%%
\begin{figure}[t]
\includegraphics[scale=0.3]{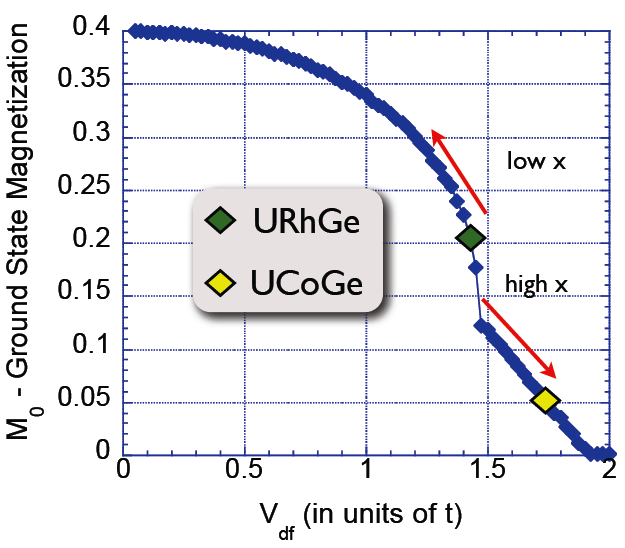}
\caption{$M_0$ from Eqs.\ (\ref{Self-Consistent-Eq}). For 
increasing $V_{df}$, a clear kink, separating the weakly and
strongly hybridized regimes, occurs at around $V_{df}\approx 1.45$,
and is also observed experimentally \cite{Huy-A-Coefficient}. } 
\label{Magnetization}
\end{figure}
%%%%%%%%%%%%%%%%%%%%%%%%%%%%%%%%%%%%%%%%%%%%%%%%%%%%%%%%%%
%

The stabilization of weakly-itinerant ferromagnetism in the periodic 
Anderson model has been studied in \cite{Batista}
%. The Hamiltonian is given by
%
\bea
H&=&-t\sum_{\langle i,j \rangle}(d_{i\sigma}^\dag d_{j\sigma}+h.c.)+
\varepsilon^0_d\sum_{i}n_{i\sigma}^{d}+\varepsilon^0_f\sum_{i}n_{i\sigma}^{f}\nonumber\\
&+&V_{df}\sum_{i}(f_{i\sigma}^\dag d_{i\sigma}+d_{i\sigma}^\dag f_{i\sigma})+
U\sum_{i}n_{i\uparrow}^{f}n_{i\downarrow}^{f},
\eea
where $t$ is the nearest neighbor hopping integral for the dispersing $d$-band, 
$\varepsilon^0_{d,f}$ are the position of the $d,f$-states, $V_{df}$ 
is the hibridization matrix element between the two, and $U$ is the on-site 
Coulomb repulsion for the $f$-level, with $n_{i\sigma}^{f}=f_{i\sigma}^\dag f_{i\sigma}$.
For large enough $U\gg V_{df}$ a Hartree-Fock mean field decoupling can be justified,
$\sum_{i}n_{i\uparrow}^{f}n_{i\downarrow}^{f}=
\sum_{i}\langle n_{i\uparrow}^{f}\rangle n_{i\downarrow}^{f}+
\sum_{i}n_{i\uparrow}^{f}\langle n_{i\downarrow}^{f}\rangle-
\sum_{i}\langle n_{i\uparrow}^{f}\rangle\langle n_{i\downarrow}^{f}\rangle$,
where $\langle n_{i\sigma}^{f} \rangle$ gives the average
occupation for the $f$ level at the site $i$, including the two 
possible spin projections. We now introduce two Hubbard Stratonovich 
fields, $N_i$ and $M_i$, associated, respectively, with the total number of
$f$ electrons, and the $f$ electron magnetization, per site,
$N_i=\langle n_{i\uparrow}^{f}\rangle+\langle n_{i\downarrow}^{f}\rangle$,
and $M_i=\langle n_{i\uparrow}^{f}\rangle-\langle n_{i\downarrow}^{f}\rangle$.
We assume the system as homogeneous and replace
$N_i=n_f=N_f/N$, where $N_f$ is the number of $f$ 
electrons and $N$ is the total number of sites. 
After this assumption we end up with
\bea
H&=&\sum_{\bf k}\varepsilon_d({\bf k})d_{{\bf k},\sigma}^\dag d_{{\bf k},\sigma}+
\overline{\varepsilon}_f\sum_{{\bf k}^\prime}f_{{\bf k}^\prime,\sigma}^\dag
f_{{\bf k}^\prime,\sigma}
+\frac{U}{4}\sum_{i}M_i^2\nonumber\\
&+&
V_{df}\sum_{{\bf k}}(f_{{\bf k},\sigma}^\dag d_{{\bf k},\sigma}+d_{{\bf k},\sigma}^\dag f_{{\bf k},\sigma})
-\frac{U}{2}\sum_{i}M_i(n_{i\uparrow}^{f}-n_{i\downarrow}^{f}),\nonumber
\eea
where
$\varepsilon_c({\bf k})=\varepsilon^0_d-2t\sum_{i=1}^{d}\cos{(k_i a)}$ is the
dispersion, and
the renormalized $f$-level is shifted to $\overline{\varepsilon}_f = \varepsilon_f+Un_f/2$.
%The term $(U/4)\sum_{i}M_i^2$ is the energy to magnetize the system, 
%while the coupling between $M_i$ and $n_{i\uparrow}^{f}$
%and $n_{i\downarrow}^{f}$ has the form of a local magnetic field. 

We introduce ferromagnetic order by writing $M_i=M_0$. The Hamiltonian 
becomes $H=N U M_0^2/4+\sum_{{\bf k}}{\cal C}^T{\cal H}{\cal C}$,
where we have defined ${\cal C}^T=(c_{{\bf k},\uparrow}^\dag c_{{\bf k},\downarrow}^\dag f_{{\bf k},\uparrow}^\dag f_{{\bf k},\downarrow}^\dag)$
and 
\beq
{\cal H}=
\left[ \begin{array}{cccc}
\varepsilon_c({\bf k}) & 0 & V_{df} & 0 \\
0 & \varepsilon_c({\bf k}) & 0 & V_{df} \\
V_{df} & 0 & \overline{\varepsilon}_f - h & 0 \\
0 & V_{df} & 0 & \overline{\varepsilon}_f + h
\end{array} \right].
\eeq
The above Hamiltonian is quadratic in the fermion operators 
and can be easily diagonalized. The new eigenvalues are 
($l=lower$,$u=upper$ branches)
{\small
\bea
E_{u,\pm}({\bf k})&=&\frac{1}{2}\left[(\varepsilon_c({\bf k})+\overline{\varepsilon}_f \pm h)+
\sqrt{(\varepsilon_c({\bf k})-\overline{\varepsilon}_f \mp h)^2+4V_{df}^2}\right],\nonumber\\
E_{l,\pm}({\bf k})&=&\frac{1}{2}\left[(\varepsilon_c({\bf k})+\overline{\varepsilon}_f \pm h)-
\sqrt{(\varepsilon_c({\bf k})- \overline{\varepsilon}_f \mp h)^2+4V_{df}^2}\right],\nonumber
\eea
}
where $h=UM_0/2$. The self-consistency equations for 
$M_0$ and $n_f$ are given by
$M_0=(1/N)\sum_{\bf k}\{\langle n_{{\bf
    k}\uparrow}^{f}\rangle-\langle n_{{\bf k}\downarrow}^{f}\rangle\}$
and $n_f=(1/N)\sum_{\bf k} \{ \langle n_{{\bf
    k}\uparrow}^{f}\rangle+\langle n_{{\bf k}\downarrow}^{f}\rangle\}$
or explicitly as
{
\begin{widetext}
\bea
M_0&=&\frac{1}{N}\sum_{\bf k}\sum_{\sigma=\pm}\mbox{sign}{(\sigma)}
\left[
\frac{E_{u,\sigma}({\bf k})-\varepsilon_c({\bf k})}{E_{u,\sigma}({\bf k})-E_{l,\sigma}({\bf k})}\; n_{FD}(E_{u,\sigma}({\bf k})-\mu)+
\frac{E_{l,\sigma}({\bf k})-\varepsilon_c({\bf k})}{E_{l,\sigma}({\bf k})-E_{u,\sigma}({\bf k})}\; n_{FD}(E_{l,\sigma}({\bf k})-\mu)
\right],\nonumber\\
n_f&=&\frac{1}{N}\sum_{\bf k}\sum_{\sigma=\pm}
\left[
\frac{E_{u,\sigma}({\bf k})-\varepsilon_c({\bf k})}{E_{u,\sigma}({\bf k})-E_{l,\sigma}({\bf k})}\; n_{FD}(E_{u,\sigma}({\bf k})-\mu)+
\frac{E_{l,\sigma}({\bf k})-\varepsilon_c({\bf k})}{E_{l,\sigma}({\bf k})-E_{u,\sigma}({\bf k})}\; n_{FD}(E_{l,\sigma}({\bf k})-\mu)
\right],
\label{Self-Consistent-Eq}
\eea
\end{widetext}
where $n_{FD}(E-\mu)=(e^{\beta(E-\mu)}+1)^{-1}$,
with $\beta=1/k_B T$. We find $M_0(x=0)=0.22\mu_B/$U-atom, for URhGe,
and $M_0(x=1)=0.06\mu_B/$U-atom, for UCoGe, in very good agreement 
with experiments \cite{Huy-A-Coefficient}, showing that in UCoGe 
$V_{df}$ is indeed much larger, see Fig.\ \ref{Magnetization}. 
We have also calculated the total density of states for the above 
band structure using Eqs.\ \ref{Self-Consistent-Eq}, with $n_f=0.4$
\cite{Gasche} and $U=9$ (in units of $t$). The results in Fig.\ \ref{Fig-DOS-PAM} 
are in very good agreement with LSDA+U \cite{Shick} and DFT data
\cite{Divis-Co}. Notice that the splitting between $+$ and $-$
states near $E_F$ is proportional to $h=UM_0/2$, being thus much 
larger for URhGe than for UCoGe, see Fig.\ \ref{Fig-DOS-PAM}. The
change in behavior of $M_0(V_{df})$ occurs exactly when $V_{df}$
becomes larger than such splitting, and $M_0$ decreases linearly
as a function of $V_{df}$, before it vanishes for 
$V_{df}^c\approx 1.9$ (in units of $t$), at the ferromagnetic QCP.

Since the 
$d$-bandwidth, $W_d$, {\it decreases monotonically} as Rh is replaced 
by Co, we parametrize
\beq
W_d(x)=W_d^{Rh}(1-x)+W_d^{Co} x,
\eeq
where the values for $W_d^{Rh}$ and $W_d^{Co}$ agree with
\cite{Divis-Rh,Divis-Co}, giving $t^{Rh}=1$ eV and 
$t^{Co}=0.6$ eV. Analogously, $\varepsilon_d^0=C_{T_d}$ 
approaches continuously to $\varepsilon_f^0=C_{U_f}$, 
with $C^{Rh}_{T_d}=-2$ eV, $C^{Rh}_{T_d}=-1.5$ eV, and 
$C_{U_f}=0$ eV, before hybridization. However, this approach, 
though monotonic, {\it deviates from linearity} \cite{Gasche} 
and thus we write
\beq
\Delta C_{df}(x)=\Delta C^{Rh}_{df}(1-x)+\Delta C^{Co}_{df} x +
\delta^\prime x^2 (1-x) + \delta^{\prime\prime} x (1-x)^2,
\eeq
where $\delta^\prime$ and $\delta^{\prime\prime}$ are adjustable 
parameters. Below we plot $T_C(x)$ for {\urhcoge} where the 
$V_{df}$ hybridization induced, non-monotonic behavior 
for $T_C$ can be observed. 
 
%
%%%%%%%%%%%%%%%%%%%%%%%%%%%%%%%%%%%%%%%%%%%%%%%%%%%%%%%%%%%
\begin{figure}[h]
\includegraphics[scale=0.26]{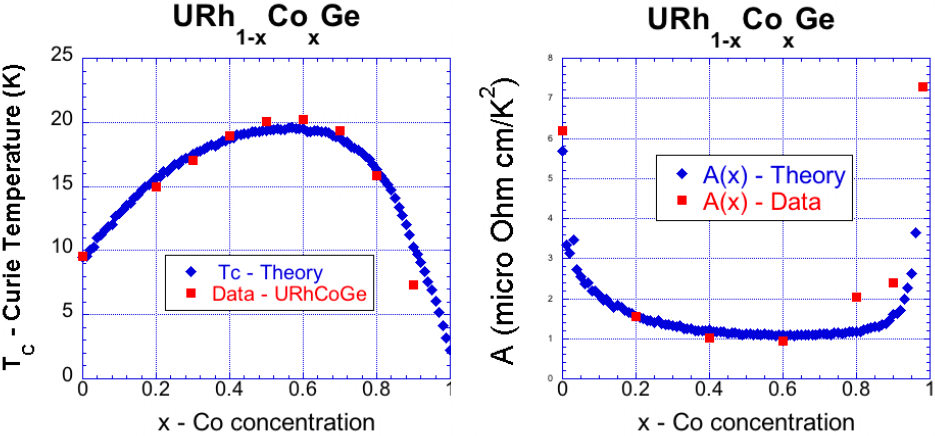}
\caption{Plot of $T_C(x)$ from Eqs.\ (\ref{Self-Consistent-Eq}) 
and of $A(x)$ from Eq.\ (\ref{A-mag}). Experimental data for 
$A$ was extracted from
\cite{Huy-A-Coefficient}. We have fixed $W_f$ so that
$T_C(x=0)=9.5$ K, in which case $V_{df}(x=0)=1.42$ and 
$V_{df}(x=1)=1.73$, see also Fig.\ \ref{Magnetization}. The 
values for the adjustable parameters, $\delta^\prime=1$ and 
$\delta^{\prime\prime}=0.8$, where used throughout the paper 
(in units of $t$). Notice the kink in $T_C(x)$, for $x=0.9$,
at the $V_{df}$ crossover, see Fig.\ \ref{Magnetization}.} 
\label{Phase-diag}
\end{figure}
%%%%%%%%%%%%%%%%%%%%%%%%%%%%%%%%%%%%%%%%%%%%%%%%%%%%%%%%%%
%

The nontrivial evolution of $N(E_F)$ governed by $V_{df}$ has also 
measurable consequences in transport properties. For example, the 
resistivity of {\urhcoge} behaves as $\rho\sim AT^2$, for all $x$, 
where $A$, the resistivity coefficient, exhibits also a nonmonotonic 
behavior as a function of $x$ \cite{Huy-A-Coefficient}. Although it 
is generally difficult to separate electronic and magnetic 
contributions to $A$, for the case of weakly-itinerant 
ferromagnets, the dominant magnetic scattering contribution to the resistivity 
is \cite{Moriya}
\beq
\rho_{mag}\sim |I N(EF)-1|^{-1/2} T^2=A T^2.
\label{A-mag}
\eeq
In fact, $A$ can be thought of as a direct measure of the proximity 
to the ferromagnetic QCP, being large for URhGe but even larger for 
UCoGe, smaller at intermediate doping $0<x<1$ \cite{Huy-A-Coefficient}. 
Thus $\rho_{mag}$ should have the opposite behavior as $T_C$, see 
Fig.\ \ref{Phase-diag}, right panel. 
%The fact that $T_C(x=0)>T_C(x=1)$,
%together with $A_{mag}(x=0)<A_{mag}(x=1)$, is in agreement with 
%values for the specific heat coefficient $\gamma(x=0)>\gamma(x=1)$.

It is interesting to compare the above results for {\urhcoge} with 
the case of another isovalent-disordered compound, {\urhgesi}. The 
substitution Ge $\rightarrow$ Si in {\urhgesi} does not modify the 
$V_{df}$ hybridization, and causes, at most, a small broadening of 
the Rh, $4d$-band, due to substitutional disorder. Such small broadening 
causes an almost negligible reduction of the $5f$-DOS at the Fermi 
level, and thus $T_C$ is expected to be {\it slightly reduced} at 
the point of largest disorder, $x=0.5$. This is exactly what we 
measured for URhGe$_{0.5}$Si$_{0.5}$ where we found $T_C(x=0.5)=8.5$ K, 
only $10 \%$ smaller than the endpoints $T_C(x=0)=T_C(x=1)=9.5$ K.

We have shown that in nearly critical, weakly itinerant ferromagnets, 
such as {\urhcoge}, the proximity to a ferromagnetic QCP is controlled 
by the hybridization, $V_{df}$, between the transition metal and the 
actinide. Smaller values for $V_{df}$ drive 
the system away from the QCP, while larger values for $V_{df}$ pushes 
the system towards the QCP. The unusual behavior observed for $T_C(x)$, 
$M_0(x)$, and $A(x)$, can all be quantitatively understood from the 
nontrivial evolution of $V_{df}(x)$ as the $4d$-band of Rh changes 
character to the $3d$-band of Co.

A.H.C.N. acknowledges DOE grant DE-FG02-08ER46512 and ONR 
grant MURI N00014-09-1-1063. Work at Florida supported by the US 
DOE, grant no. DE-FG02-86ER45268.

\end{document}